\documentclass[prc,preprint,showpacs,showkeys,nofootinbib]{revtex4}%
\usepackage{graphicx}
\usepackage{bm}
\usepackage{epsfig}
\usepackage{amsmath}
\usepackage{amsfonts}
\usepackage{amssymb}%
\begin{document}

\title{Electronic stopping in astrophysical fusion reactions}
\author {C.A. Bertulani \footnote{e-mail: bertulani@nscl.msu.edu}}
\affiliation{National Superconducting Cyclotron Laboratory,
Michigan State University, East Lansing, MI 48824, USA}
\date{\today}

\begin{abstract}
The stopping power of protons and deuterons in low energy
collisions with helium gas targets is investigated with the
numerical solution  of the time-dependent Schr\"odinger
coupled-channels equations using molecular orbital wavefunctions.
It is shown that at low projectile energies the energy loss is
mainly due to nuclear stopping, charge exchange, and the
excitation of low energy levels in the target. The second and
third mechanisms, called electronic stopping, dominate for
E$_{lab}\geq200$ eV. At lower energies it is also shown that a
threshold effect is responsible for a quick drop of the energy
loss. This investigation sheds more light on the long standing
electron screening problem in fusion reactions of astrophysical
interest.

\end{abstract}

\maketitle


\narrowtext

Nuclear fusion reactions proceed in stars at extremely low
energies, e.g. of the order of $10$ keV in our sun
\cite{Cla68,RR88}. At such low energies it is extremely difficult
to measure the cross sections for charged particles at laboratory
conditions due to the large Coulomb barrier. Moreover, laboratory
measurements of low energy fusion reactions are strongly
influenced by the presence of the atomic electrons. One has
observed experimentally a large discrepancy between the
experimental data and the best models available to treat the
screening effect by the electrons in the target nuclei
\cite{Ro01}. The screening effect arises because as the projectile
nucleus penetrates the electronic cloud of the target the
electrons become more bound and the projectile energy increases by
energy conservation. Since fusion cross sections increase strongly
with the projectile's energy, this tiny amount of energy gain (of
order of 10-100 eV) leads to a large effect on the measured cross
sections. However, in order to explain the experimental data it is
necessary an extra-amount of energy about twice the expected
theoretical value \cite{Ro01}.

In order to extract the fusion cross sections from experiment one
needs to correct for the energy loss in the target to assign the
correct projectile energy value for the reaction. The authors in
refs. \cite{LSBR96} and \cite{BFMH96} have shown that a possible
solution to the long standing discrepancy between theory and
experiment for the reaction $^{3}$He(d$,$ p)$^{4}$He could be
obtained if the projectile energy loss by electronic excitations
and charge exchange with the target atoms would be smaller than
previously assumed in the experimental data analysis. There have
been indeed a few experiments in which evidences of smaller than
expected electronic stopping power were reported (see, e.g. ref.
\cite{GS91}). Other reactions of astrophysical interest (e.g.,
those listed in by Rolfs and collaborators \cite{ALR87,SR95})
should also be corrected for this effect. Whereas at higher
energies the stopping is mainly due to the ionization of the
target electrons, at the astrophysical energies it is mainly due
to excitations of the lowest levels, charge-exchange between the
target and the projectile, and the nuclear stopping power.

In this work I address the problem of the stopping of very low
energy ions in matter. I consider the systems p+$^{4}$He and
d+$^{3}$He, for which there are experimental data available. A
previous work \cite{BD00} studied the energy loss of protons on
hydrogen gas targets and showed that the stopping at very low
proton energies is indeed smaller than what would be expected from
extrapolations based on the Andersen and Ziegler tables
\cite{AZ77}. The case of helium targets is more complicated due to
the electron-electron interaction.

The present approach is based on the solution of the time-dependent
Schr\"{o}dinger equation for the electron in a dynamical two-center field. The
transition from the separated atoms ($\mathrm{H}^{+}+\mathrm{He}$) and the
united atom (Li$^{+}$) is obtained in the adiabatic approximation, i.e. by
assuming that the electronic motion is fast compared to the nuclear separation
motion so that the molecular orbitals (MO) are those for the distance $R(t$)
between the nuclei. The atomic wavefunctions, $\phi_{\mu}=\sum_{j}c_{j\mu}%
\phi_{j}^{Slat}$, are constructed as a linear combination of Slater-type
orbitals (STO)  \cite{Lev00} of the form $\phi_{n}^{Slat}=Nr^{n-1}%
\exp\left(  -\zeta r\right)  Y_{lm}\left(
\widehat{\mathbf{r}}\right)  $, where the Slater coefficients $n$
and $\zeta$ are chosen to best approximate the exact atomic
wavefunctions (see, e.g. ref. \cite{Lev00}). The molecular orbital
wavefunctions for the $\mathrm{p}+\mathrm{He}$ system, are
obtained with the $\phi_{\mu}$'s chosen so that half of the STO's
are centered on the proton $(A)$ and the other half are centered
on the helium nucleus $(B)$. The total wavefunction for the
two-electron system is finally written as a Slater
determinant of the molecular orbital wavefunctions,%
\begin{equation}
\psi_{e}\left(  \mathbf{r}_{1},\mathbf{r}_{2},R\right)  =\frac{1}{\sqrt{2}%
}\left\vert
\begin{array}
[c]{cc}%
\phi_{1}^{MO}(1)\alpha\left(  1\right)  & \phi_{2}^{MO}(1)\beta\left(
1\right) \\
\phi_{1}^{MO}(2)\alpha\left(  2\right)  & \phi_{2}^{MO}(2)\beta\left(
2\right)
\end{array}
\right\vert ,\label{phielec}%
\end{equation}
where $\alpha,\beta$\ denote the spin state of the electron.
Configuration-interaction with double excitation configurations were included
in the calculation, with the coefficients $n$ and the Slater parameters
$\zeta$\ chosen in a variational method to obtain the lowest energy states of
the system.

Using these conditions and variation method, one obtains the following
Hatree-Fock equation: $\mathbf{F}\otimes\mathbf{C=O}\otimes\mathbf{C}%
\otimes\mathbf{E} $, where \textbf{F} is the \textquotedblleft Fock" matrix%
\begin{equation}
F_{\mu\nu}=H_{\mu\nu}+\sum_{\lambda\sigma}P_{\lambda\sigma}\left[
\left\langle \mu\nu\left\vert \frac{1}{r_{12}}\right\vert \lambda
\sigma\right\rangle -\frac{1}{2}\left\langle \mu\lambda\left\vert
\frac {1}{r_{12}}\right\vert \nu\sigma\right\rangle \right] ,\ \ \
\ \ \ \ \ \ \ \ P_{\lambda\sigma}=2\sum_{i=1}^{occ}c_{\lambda
i}c_{\sigma i}\ ,
\end{equation}
in which \textquotedblleft occ" refers to the occupied molecular orbital,%
\begin{equation}
H_{\mu\nu}=\int\int \phi_{\mu} \left(  1\right)  \left[
-\frac{1}{2}\nabla ^{2}-\sum_{L=A,B}\frac{1}{r_{1L}}\right]
\phi_{\nu}\left( 1\right) d\tau_{1}\ ,
\end{equation}
is the one-electron integral and%
\begin{equation}
\left\langle \mu\nu\left\vert \frac{1}{r_{12}}\right\vert \lambda
\sigma\right\rangle =\int\int\phi_{\mu}\left(  1\right)  \phi_{\nu}\left(
1\right)  \frac{1}{r_{12}}\phi_{\lambda}\left(  2\right)  \phi_{\sigma}\left(
2\right)  d\tau_{1}d\tau_{2}\ ,
\end{equation}
are the two-electron integrals. The \textbf{C} matrix is the coefficient
matrix $c_{\mu\nu}$ and \textbf{O} is the overlap matrix $\left\langle
\phi_{\mu}\left(  1\right)  |\phi_{\nu}\left(  1\right)  \right\rangle
.$\ \textbf{E} is a diagonal matrix with each diagonal element corresponding
to the energy of the associated molecular orbital. Solving the Hartree-Fock
equations one obtains the coefficients $c_{ij}$ which give the proper linear
combination of atomic orbitals to form the molecular orbital. The energy of
the molecular orbitals are then given by%
\begin{equation}
E(R)=\sum_{\mu\nu}P_{\mu\nu}H_{\mu\nu}+\frac{1}{2}\sum_{\mu\nu\lambda\sigma
}P_{\mu\nu}P_{\lambda\sigma}\left[  \left\langle \mu\nu\left\vert \frac
{1}{r_{12}}\right\vert \lambda\sigma\right\rangle -\frac{1}{2}\left\langle
\mu\lambda\left\vert \frac{1}{r_{12}}\right\vert \nu\sigma\right\rangle
\right]  \ .
\end{equation}

\begin{figure}
[ptb]
\begin{center}
\includegraphics[
height=3.4in, width=2.9in
]%
{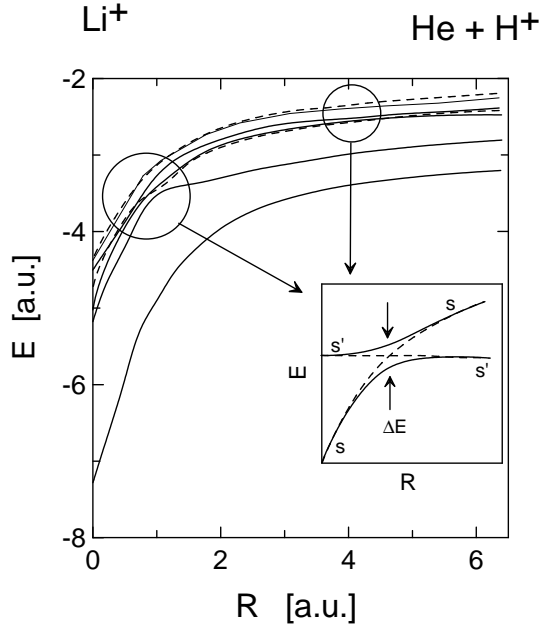}%
\caption{Adiabatic energies (1 a.u. of energy = 27.2 eV, 1 a.u. of
length = 0.53 \AA ) for the electronic orbitals for the
(H-He)$^{+}$ system as a function of the internuclear separation.
As the atoms approach each other slowly curves of same symmetry
repel each other. A transition between states s and s' can occur
in a slow collision. In a fast collision a diabatic transition,
with the states crossing each other, will occur. This is shown in
the inset.}%
\label{f1}%
\end{center}
\end{figure}

Table I shows the states involved in the calculation where it is
shown how the states in the separated hydrogen and helium atoms
become molecular states in the united atom system. For large
distances between the nuclei, $R>15$ a.u. (1 a.u. of length = 0.53
\AA ) the energy levels for the 1s, 2s, and 2p states of H and He
are reproduced to within 2\% and 4\% of the spectroscopic data,
respectively. The energies of these states are shown in figure
\ref{f1} as a function of the internuclear distance $R$.

\begin{center}%
\begin{tabular}
[c]{|l|l|}\hline
Separated atom & United atom\\\hline
$\mathrm{H}^{+}+\mathrm{He}(1s^{2})$ & 0$\Sigma$\\\hline
$\mathrm{H}(1s)+\mathrm{He}^{+}(1s)$ & 1$\Sigma$\\\hline
$\mathrm{H}^{+}(1s)+\mathrm{He}(1s2s)$ & 2$\Sigma$\\\hline
$\mathrm{H}(n=2)+\mathrm{He}^{+}(1s)$ & 1$\Pi$\\\hline
$\mathrm{H}(n=2)+\mathrm{He}^{+}(1s)$ & 3$\Sigma$\\\hline
$\mathrm{H}(n=2)+\mathrm{He}^{+}(1s)$ & 4$\Sigma$\\\hline
$\mathrm{H}^{+}+\mathrm{He}(1s1p)$ & 5$\Sigma$\\\hline
$\mathrm{H}^{+}+\mathrm{He}(1s1p)$ & 2$\Pi$\\\hline
\end{tabular}

Table 1 - Lowest states in the p+He molecule.
\end{center}

At very low proton energies ($E_{P}\lesssim10$ \textrm{keV}) it is
fair to assume that only the low-lying states are involved in the
electronic dynamics. Only for bombarding energies larger than 25
keV the proton velocity will be comparable to the electron
velocity, v$_{e}\simeq\alpha c$. Thus, the evolution of the system
is almost adiabatic at $E_{p}\lesssim10$ keV.  Also shown in
figure \ref{f1} (inset) are the intersection points of the states
with same symmetry. In a fast collision these states would cross
(diabatic collisions), whereas in collisions at very low energies
(adiabatic collisions) they obey the von Neumann-Wigner
non-crossing rule.

In the dynamical case the full time-dependent wavefunction for the system can
be expanded in terms of two-center states, $\psi_{n}\left(  \mathbf{r}%
_{1},\mathbf{r}_{2},t\right)  $, given by eq. \ref{phielec},\ with expansion
coefficients $a_{n}\left(  t\right)  $. It is further assumed that the nuclei
follow a classical straight-line trajectory determined by an impact parameter
$b$, so that the time dependence of the molecular wavefunctions is determined
by the condition $R=\sqrt{b^{2}+v^{2}t^{2}}$, where $v$ is the collision
velocity. The dynamical evolution of the H+He system is calculated using the
same approach as described in ref. \cite{BD00}. We solve the set of linear
coupled equations%
\begin{equation}
i\mathbf{S\cdot}\frac{d\mathbf{A}}{dt}\mathbf{=M\cdot A}\label{cc}%
\end{equation}
where the column matrix \textbf{A} represents the time-dependent expansion
coefficients, \textbf{S} is the overlap matrix with elements $S_{ij}%
=\left\langle \psi_{i}|\psi_{j}\right\rangle $ and \textbf{M} is the coupling
matrix with elements $M_{ij}=\left\langle \psi_{i}\left\vert H_{el}%
-i\partial/\partial t\right\vert \psi_{j}\right\rangle $, where $H_{el}$\ is
the electronic Hamiltonian. The solutions are obtained starting from initial
internuclear distance of 15 a.u. for the incoming trajectory and stopped at
the same value for the outgoing trajectory. The probability for the capture in
the proton is obtained by a projection of the final wavefunction into the
wavefunctions of \ the $1s$, $2s$ and and $2p$ states of the hydrogen atom.%
\begin{equation}
P_{exch}=\left\vert \sum_{m}a_{m}\left(  \infty\right)  \left\langle \psi
_{H}|\psi_{m}\left(  \infty\right)  \right\rangle \right\vert ^{2}%
\;.\label{approx}%
\end{equation}

Resonant charge-exchange in atomic collisions was first observed
by Everhart and collaborators \cite{Ever}. In these experiments it
was determined that the exchange probability in homonuclear atomic
collisions oscillates with the incoming energy for a collision
with a given impact parameter, or scattering angle. This was
interpreted \cite{Hol52} as due to transitions between degenerate
states of the system at large internuclear separation distance. In
the simplest situation of a $\mathrm{p}+\mathrm{H}$ the degenerate
states are the symmetric and antisymmetric states obtained from
the linear combination of the (H$^{+}$H) and (HH$^{+}$)
wavefunctions. This effect was studied in ref. \cite{Lich63},
where a relation between the damping of the oscillatory behavior
of the exchange probabilities and the Landau-Zener effect was
established. The $\mathrm{p}+\mathrm{H}$ collisions at small
energies was recently studied in ref. \cite{BD00} and the
oscillatory effect was shown to be related to the Sommerfeld
quantization rule for the integral from $t=-\infty$ to
$t=-\infty$\ of the energy difference between the symmetric and
antisymmetric state. The electron tunnels back and forth between
the projectile and the target during the ingoing and the outgoing
parts of the trajectory. When the interaction time is an exact
multiple of the oscillation
time, a minimum in the exchange probability occurs.%
\begin{figure}
[ptb]
\begin{center}
\includegraphics[
height=2.9594in,
width=2.9827in
]%
{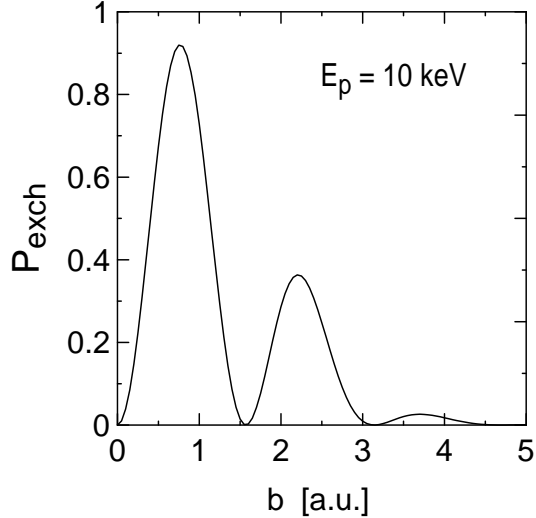}%
\caption{Probability of charge exchange in the collision p+$^4$He
showing the resonant behavior as a function of the impact
parameter and for proton energy $E_p=10$ KeV.} \label{f2}
\end{center}
\end{figure}

A similar situation occurs for $\mathrm{p}+\mathrm{He}$
collisions, as shown in figure \ref{f2} for the electron capture
probability by the proton at 10 keV bombarding energy. These
oscillations are due to the electron exchange between the ground
state of the hydrogen and the first excited state in He (1s2s).
But, in contrast to the H$^{+}$H system, the oscillations are
strongly damped. Following the work of Lichten \cite{Lich63}\ we
interpret this damping effect as due to the interference between
the participant states and a band of states of average energy
$\left\langle E_{a}\right\rangle $\ and width 2$\Gamma$, as seen
in figure \ref{f1}. The important regions where the diabatic level
cross occurs is shown in figure \ref{f1} inside the encircled
areas. The damping mechanism is best understood using the
Landau-Zener theory for level crossing. At the crossing there is a
particular probability ($1-P$) of an adiabatic transition where
$P$ is given by the
Landau-Zener formula%
\begin{equation}
P_{exch}=\exp\left[  \frac{2\pi H_{ss^{\prime}}^{2}}{v\left(  d/dR\right)
\left(  E_{s}-E_{s^{\prime}}\right)  }\right] \label{lz}%
\end{equation}
where $v$ is the collision velocity and $H_{ss^{\prime}}$ is the
off-diagonal matrix element connecting states $s$ and $s\prime$.
The oscillatory behavior shown in figure \ref{f2} is due to the
many level transitions at the crossing, each time governed by the
probability of eq. \ref{lz}. The interference with the neighboring
states introduces a damping in the charge exchange probability,
i.e.%
\[
P_{exch}\left(  b,t\right)  \simeq\cos^{2}\left(  \frac{\left\langle
E_{a}\right\rangle b}{v}\right)  \exp\left[  -\frac{2\pi\Gamma^{2}%
b}{v\left\langle E_{a}\right\rangle }\right]  ,
\]
where $\left\langle E_{a}\right\rangle \simeq1.$ a.u. is the
average separation energy between the 0$\Sigma$ level and the
bunch of higher energy levels shown in figure \ref{f1}. The
exponential damping factor agrees with the numerical calculations
if one uses $\Gamma\simeq5$ eV, which agrees with the energy
interval of the
band of states shown in figure \ref{f1}.%
\begin{figure}
[ptb]
\begin{center}
\includegraphics[
height=2.7in, width=3.in ] {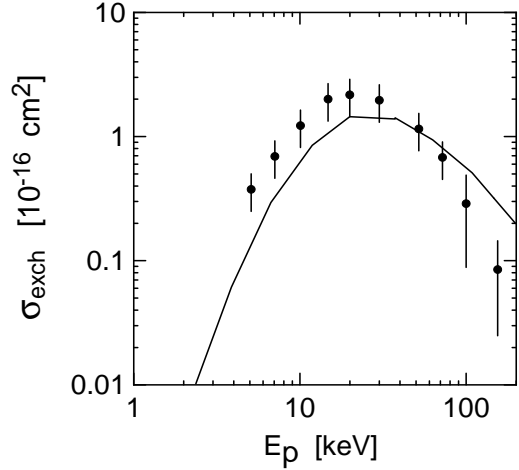} \caption{Charge-exchange
cross sections for the $\mathrm{p}+ ^4\mathrm{He}$ as a function
of the proton energy. The solid line was obtained by solving the
coupled-channels equations \ref{cc}, and using \ref{approx} for
the exchange probability. The experimental data are from ref.
\cite{Rud83}.} \label{phexsec}
\end{center}
\end{figure}

The total cross section for charge exchange is calculated from
\[
\sigma=2\pi\int P_{exch}\ bdb\ .
\]
The numerical results for the $\mathrm{p}+\mathrm{He}$ system is
shown in figure \ref{phexsec} as a function of the proton energy.
The solid line is the result of using the coupled-channels
equations \ref{cc}, and \ref{approx} for the exchange probability.
The experimental data are from ref. \cite{Rud83}. We observe that
the calculation reproduces the trend of the experimental data.
But, whereas the maximum of the cross section at $E_{p}\simeq20$
keV is rather well described, the calculations underestimate the
cross sections at smaller energies. The low energy slope of the
cross section is nonetheless well reproduced. At energies higher
than the Bragg peak ($E_{p}\gtrsim20$ keV) the numerical results
should not be trusted as the adiabatic approximation for the
molecular orbitals and also the inclusion of only the lowest
energy levels are not adequate (continuum states should also be
included). For $E_{p}\longrightarrow0$, the charge exchange cross
section must go to zero since the higher binding of the electrons
in He prevents the capture by the incident proton in an extreme
adiabatic collision. This feature
is correctly reproduced by the numerical calculations.%
\begin{figure}
[ptb]
\begin{center}
\includegraphics[
height=3.1272in,
width=2.8591in
]%
{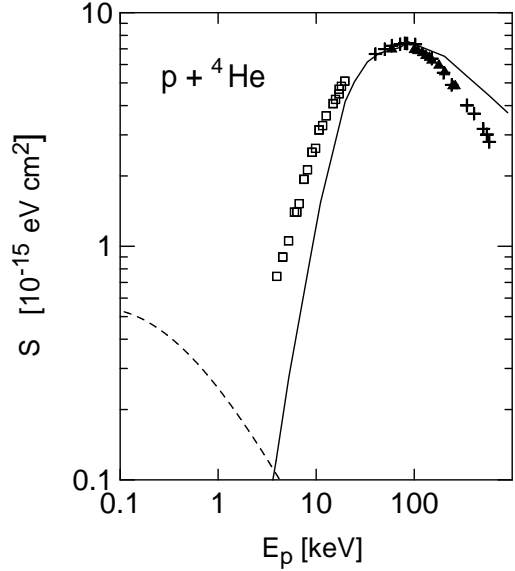} \caption{Stopping cross section for proton incident on
gas $^4$He targets, as a function of the proton energy. The
experimental data are from refs. \cite{Rud83,PZ63}.}
\label{stopHe}%
\end{center}
\end{figure}

In figure \ref{stopHe} we show the stopping cross section of the
proton. The experimental data are from refs. \cite{Rud83,PZ63}.
The stopping cross section is defined as $S=\sum_{i}\Delta
E_{i}\;\sigma_{i}$ , where $\Delta E_{i}$ is the energy loss of
the projectile in a process denoted by $i$. The stopping power,
$S_{P}=dE/dx$, the energy loss per unit length of the target
material, is related to the stopping cross section by $S=S_{P}/N$,
where $N$ is the atomic density of the material. In the charge
exchange mechanism one of the electrons in He is transferred to
incoming proton and the energy loss by the proton is given by
$\Delta E=m_{e}v^{2}/2$, where $v$ is the proton velocity.
Assuming that there are few free electrons in the material (e.g.
in the helium gas) only one more stopping mechanism at very low
energies should be considered: the nuclear stopping power,
$S_{n}$. This is simply the elastic scattering of the projectile
off the target nuclei. The projectile energy is partially
transferred to the recoil energy of the target atom. The stopping
cross section for this mechanism has been extensively discussed in
ref. \cite{ZBL85}. The total stopping power is given by
$S=S_{exch}+S_{n}$.\ In units of $10^{-5}$ eV cm$^{2}$ the nuclear
stopping for the p+$^{4}$He system
at $E_{p}<30$ keV is given by%
\begin{equation}
S_{n}=S_{0}\ \frac{\ln\left(  1+1.1383\varepsilon\right)  }{\left(
\varepsilon+0.01321\varepsilon^{0.21226}+0.19593\varepsilon^{0.5}\right)
}\ ,\label{nstop}%
\end{equation}
where $S_{0}=0.779$ and $\varepsilon=5.99E_{p},$ with the proton incident
energy $E_{p}$ given in keV.

The dashed line in figure \ref{stopHe} gives the energy transfer by means of
nuclear stopping, while the solid line shows the results for the electronic
stopping mechanism, i.e. due \ to charge-exchange and excitation in the helium
target. We see that the nuclear stopping dominates at the lowest energies,
while the electronic stopping is larger for proton energies greater than 200
eV. We do not consider the change of the charge state of the protons as they
penetrate the target material. The exchange mechanism transforms the protons
into H atoms. These again interact with the target atoms. They can loose their
electron again by transferring it back to a bound state in the target.

At very low energies the only possibility that the electron is
captured by the proton is if there is a transition
1s$^{2}$($^{1}$S$_{0}$) $\longrightarrow$ 1s2s($^{3}$S) in the
helium target. Only in this case the energy of one of the
electrons in helium roughly matches the electronic energy of the
ground state in H. This resonant transfer effect is responsible
for the large capture cross sections. When this transition is not
possible the electrons prefer to stay in the helium target, as the
energy of the whole system is lowest in this case. Another
possible mechanism for the stopping is the excitation of the
helium atom by the transition 1s$^{2}$($^{1}$S$_{0}$)
$\longrightarrow$ 1s2s($^{3}$S). Thus, there must be a direct
relationship between the energy transfer to the transition
1s$^{2}$($^{1}$S$_{0}$) $\longrightarrow$ 1s2s($^{3}$S) and the
minimum projectile energy which enables electronic changes. Ref.
\cite{For00} reported for the first time this effect, named by
threshold energy, which can be understood as follows. The momentum
transfer in the projectile-target collision, $\Delta q$, is
related to the energy transfer to the electrons by $\Delta
q=\Delta E/v$, where $v$ is the projectile velocity. In order that
this momentum transfer absorbed by the electron, induces an atomic
transition, it is necessary that $\hbar^{2}\Delta
q^{2}/2m_{e}\sim\Delta E$. Solving these equations for the
projectile energy one finds
\begin{equation}
E_{p}^{thres}\sim\frac{m_{p}}{4m_{e}}\Delta E\ .\label{ethres}%
\end{equation}
This is the threshold energy for atomic excitations and/or charge exchange. If
the projectile energy is smaller than this value, no stopping should occur.
The energy for transition 1s$^{2}$($^{1}$S$_{0}$) $\longrightarrow$
1s2s($^{3}$S) in He is $\Delta E=18.7$ eV. Thus, for $\mathrm{p}+\mathrm{He}$
collisions, the threshold energy is $E_{p}^{thres}\sim9$ keV. This roughly
agrees with the numerical calculations presented in figure \ref{stopHe} (solid
curve).%
\begin{figure}
[ptb]
\begin{center}
\includegraphics[
height=3.3546in,
width=3.0225in
]%
{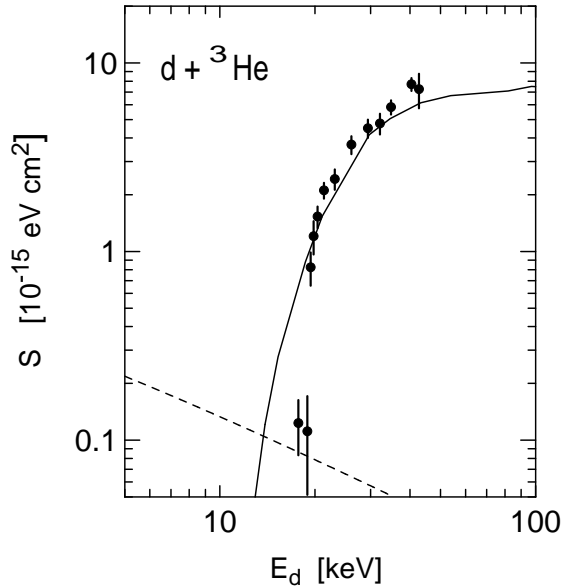} \caption{Energy loss of deuterons in $^{3}\mathrm{He}$
gas as a function of deuteron energy. Data are from  ref.
\cite{For00}. The solid curve is the calculation for the
electronic stopping power, while the dashed curve shows the
nuclear stopping.}
\label{stopHe3}%
\end{center}
\end{figure}

Figure \ref{stopHe3} shows the energy loss of deuterons in
$^{3}\mathrm{He}$ gas as a function of deuteron energy. The data
are from ref. \cite{For00}. The solid curve is  the numerical
calculation for the electronic stopping power, while the dashed
curve shows the nuclear stopping. For this system the coefficients
in eq. \ref{nstop} are $S_{0}=1.557$ and $\varepsilon=4.491E_{d},$
respectively.  As discussed in ref \ \cite{For00} the threshold
deuteron energy in this reaction is of the order of 18 keV, which
agrees with the estimate based on eq. \ref{ethres}. However, the
numerical calculations based on the electronic stopping (solid
curve of fig. \ref{stopHe3}) indicate a lower threshold energy for
this system. Nonetheless, the agreement with the experimental data
is very good for $E_{d}>20$ keV. The threshold effect is one more
indication that the extrapolation $S\sim v$, based on the
Andersen-Ziegler tables is not applicable to very low energies.

The steep rise of the fusion cross sections at astrophysical
energies amplifies all effects leading to a slight modification of
the projectile energy \cite{BBH97}. The results presented in this
article show that the stopping mechanism does not follow a
universal pattern for all systems. The threshold effect reported
in ref. \cite{For00} is indeed responsible for a rapid decrease of
the electronic stopping at low energies. It will occur whenever
the charge-exchange mechanism and the excitation of the first
electronic state in the target involve approximately the same
energy. However, the drop of the electronic stopping is not as
sharp as expected from the simple classical arguments given by eq.
\ref{ethres}.

The experiments on astrophysical fusion reactions have shown that
the screening effect is much larger than expected by theory. The
solution to this problem might be indeed the smaller stopping
power, due to a steeper slope  at low energies induced, e.g. by
the threshold mechanism. This calls for improved theoretical
studies of the energy loss of ions at extremely low energies of
and for their independent experimental verification. The present
situation is highly disturbing because if we cannot explain the
laboratory screening effect, most likely we cannot explain it in
stellar environments.

\bigskip

I would like to express my gratitude to D.T. de Paula and I.
Ivanov for helping me with the programming during the earlier
stages of this work. This work was supported by the National
Science Foundation under Grants No. PHY-007091 and PHY-00-70818.

\end{document}